\documentclass[twoside]{book}
\usepackage{graphicx}		

 \makeatletter
\ifx\UNDEF\mail\def\mail{ }\else\fi
\ifx\UNDEF\prange\def\prange{0 0}\else\fi

\gdef\@empty{}
\def\Mail#1 #2 {\gdef\thecontact{#1}\gdef\theaddr{#2}}
\def\Range#1 #2 {\gdef\thefirstpage{#1}\gdef\thelastpage{#2}}
{\let\'\mail \expandafter\Mail\' }	
{\let\'\prange \expandafter\Range\' }	
 \gdef\@shtitle{\relax}
 \long\def\shtitle#1{\gdef\@shtitle{#1}}
 \long\def\author#1{\gdef\@author{#1}}
 \def\affil#1{\par\noindent{\rm#1\par}}
 \gdef\@abstract{}
 \long\def\abstract#1{\gdef\@abstract{#1}}

 \renewcommand{\@evenhead}{\thepage\qquad\qquad\@shtitle\hfil}
 \renewcommand{\@oddhead}{\hfil\@shtitle\qquad\qquad\thepage}

 \def\maketitle{\thispagestyle{empty}\chapter{\@title}}
 \renewcommand\chapter{\if@openright\cleardoublepage\else\clearpage\fi
                    \thispagestyle{empty}%
                    \global\@topnum\z@
                    \@afterindentfalse
                    \secdef\@chapter\@schapter}
 \def\@makechapterhead#1{%
  \vspace*{50\p@}%
  {\parindent \z@ \raggedleft \normalfont
    \ifnum \c@secnumdepth >\m@ne
      \if@mainmatter
        \huge \@chapapp{} \thechapter
        \par\nobreak
        \vskip 20\p@
      \fi
    \fi
    \interlinepenalty\@M
    \Huge \bfseries #1\par\nobreak
    \vskip.25in
    \large\bfseries\@author\par\nobreak
    \vskip 40\p@}
    \ifx\@abstract\@empty\else{\small\@abstract\par\vskip20\p@}\fi
  }

\DeclareRobustCommand\em
        {\@nomath\em \ifdim \fontdimen\@ne\font >\z@
                       \upshape \else \slshape \fi}

\def\@begintheorem#1#2{\sl \trivlist \item[\hskip \labelsep{\bf #1\ #2}]}
\def\@opargbegintheorem#1#2#3{\sl \trivlist
     \item[\hskip \labelsep{\bf #1\ #2\ (#3)}]}

  \def\@arabic#1{\number #1} 

\long\def\@makecaption#1#2{
	\vskip\abovecaptionskip
	\sbox\@tempboxa{{\small {\bf #1}: #2}}%
	\ifdim\wd\@tempboxa>\hsize
	    {\small {\bf #1}: #2\par}
	\else
	   \global\@minipagefalse
	   \hbox to\hsize{\hfil\box\@tempboxa\hfil}
	\fi
	\vskip \belowcaptionskip}

\def\figstrut#1{\hbox to\linewidth{\vrule height#1\hfill}}

\renewenvironment{thebibliography}[1]
     {\section*{\bibname
        \@mkboth{\MakeUppercase\bibname}{\MakeUppercase\bibname}}%
      \list{\@biblabel{\@arabic\c@enumiv}}%
           {\settowidth\labelwidth{\@biblabel{#1}}%
            \leftmargin\labelwidth
            \advance\leftmargin\labelsep
            \@openbib@code
            \usecounter{enumiv}%
            \let\p@enumiv\@empty
            \renewcommand\theenumiv{\@arabic\c@enumiv}}%
      \sloppy
      \clubpenalty4000
      \@clubpenalty \clubpenalty
      \widowpenalty4000%
      \sfcode`\.\@m}
     {\def\@noitemerr
       {\@latex@warning{Empty `thebibliography' environment}}%
      \endlist}
\makeatother

 \shtitle{Universality in phase-ordering}	
 \title{Considerations about universality in phase-ordering of binary liquids}
 \author{Alexander J. Wagner\affil{Department of Physics and Astronomy\\
The University of Edinburgh\\awagner@ph.ed.ac.uk}}
\usepackage{psfig}

\abstract{ In this article we show that the phase-ordering scaling
state for binary fluids is not necessarily unique and that local
correlations in the initial conditions can be responsible for selecting
the scaling state. We describe a new scaling state for symmetric
volume fractions that consists of drops of the one component suspended
in a matrix of the other.  The underlying reason for the existence of
the newly observed scaling state is that the main coarsening mechanism
of binary fluids which is the deformation of interfaces by flow is not
acting, and this leads to a new scaling law. An initial droplet state
can be formed by a number of physical phenomena. In a unified
description this can be undestood as local correlations in the initial
conditions. Local correlations with length $\xi$ are believed to be
irrelevant when the typical length scale L of the system is large
($L\gg \xi$). Our result shows that these initial correlations,
contrary to current thinking, can be important even at late times.}

\begin{document}
\maketitle

\section{Introduction}
Phase-ordering is observed in many spatial systems, including ones
that underwent spinodal decomposition or mechanical mixing. Examples
include a wide variety of systems including magnetic systems, binary
alloys, binary fluids, but also some reaction diffusion
systems\cite{Epstein} and some models of gene evolution\cite{Sayama}.

In a seminal paper Hohenberg and Halperin showed that most of these
systems can be categorized to belong to a small number of categories,
and each member of a category shows the same universal phase-ordering
behavior. In particular they distinguished between systems in which
the order parameter is not conserved,{\it e.g.} magnetic systems or
genetic systems, and systems in which the order parameter is conserved,
{\it e.g.} binary alloys or binary fluids. Liquid systems, such as binary
fluids and magnetic fluids, can phase-order not only by diffusion but
also via flow and therefore they are in a universality class of their own.

The phase-ordering process has been conjectured to observe scaling,
{\it i.e.} the systems are characterized by a single length-scale
$L(t)$ such
that the morphologies of the system scaled but that length-scale are
statistically self-similar. The time dependence of this length-scale
is then known as the scaling law. The scaling laws can be deduced by
dimensional analysis from the equations of motion. For systems without
hydrodynamics $L(t)$ scales as $t^{1/2}$ for systems with
non-conserved order parameter and as $t^{1/3}$ for systems with a
conserved order parameter.

Because binary fluids have more that one mechanism for domain
coarsening there is more that one growth law. Off-critical quenches
that contain a much smaller volume fraction of one component form
droplets and this morphology grows mostly via diffusion leading to a
$L(t)\sim t^{1/3}$ growth law. Most studies, however, deal with
critical quenches with equal volume fractions for both
components. Because of this symmetry bi-continuous morphologies are
typically formed and both diffusion and hydrodynamic growth mechanisms
operate.

There are, however, exceptions to this general rule. Sometimes
dynamical asymmetries in the diffusion constant of the two phases or
their viscoelastic properties can lead to droplet morphologies.
A stationary morphology that consists of circular droplets
has no hydrodynamic pathway of coarsening. Once two droplets coalesce
they can induce a flow that can induce further coalescence.  The
central idea of this paper is that one of the main hydrodynamics
pathways of coarsening can be suppressed if the morphology consists of
droplets leading to a new scaling state. The possibility of the
non-uniqueness of the scaling state was first suggested by
A. Rutenberg, although at the time no persuasive numerical evidence
could be found\cite{Rutenberg}.  

The droplet morphologies that we require are not unusual and can often
seen in the early stages of viscoelastic spinodal decomposition
\cite{Bhattacharya,Haas,Onuki} as well as in systems with an
order-parameter dependent mobility. We will use a viscoelastic lattice
Boltzmann method to create the initial droplet morphology and then use
a symmetric Newtonian lattice Boltzmann method to evolve the initial
droplet morphology to verify that this new scaling state exists.

In viscoelastic phase-separation the droplet morphology consists of
low-viscosity droplets suspended in the viscoelastic matrix. The
opposite morphology is observed in mechanical mixing where the low
viscosity material will form the matrix phase. We show that both
droplet morphologies are stable under phase-ordering, in agreement
with experimental results.

\section{Numerical method}
For the simulations we use the viscoelastic two-component
lattice Boltzmann simulation introduced in an earlier
paper\cite{vedrop}. Briefly, in lattice Boltzmann simulations
densities $f_i$ that are associated with velocities $v_i$ are streamed
on a lattice according to the lattice Boltzmann equation
\begin{eqnarray}
&&f_i({\bf x}+{\bf v}_i \Delta t,t+\Delta t)=\nonumber\\
&&f({\bf x},t)
+ \Delta t \sum_j \Lambda_{ij} \left[f_j^0({\bf x},t)-f_j({\bf x},t)\right]
\label{eqnLB1}
\end{eqnarray}
where $f_i^0$ is an equilibrium distribution, $\Lambda_{ij}$ is a
collision matrix, and ${\bf v}_i \Delta t$ is a lattice vector. The
velocity set for our simulation consists of 17 velocities given by
$\{(0,0)$, $(1,0)$, $(0,1)$, $(-1,0)$, $(0,-1)$, $(1,1)$, $(-1,1)$,
$(-1,-1)$, $(1,-1)$, $(1,0)$, $(0,1)$, $(-1,0)$, $(0,-1)$, $(1,1)$,
$(-1,1)$, $(-1,-1)$, $(1,-1)\}$. Note that the last 8 velocities are
the same as the previous eight velocities. This duplicity allows
the simulation to have two independent stresses which represent a
viscoelastic and a purely viscous contribution to the total stress
tensor. The two contributions are used to produce a Jeffrey's model
for the stress (see eqn. (\ref{jeffrey})). The algorithm is required
to conserve mass and momentum, but not energy. Energy conservation is
replaced by a condition of constant temperature. The macroscopic
density $\rho$ and velocity ${\bf u}$ are defined as
\begin{equation}
\rho=\sum_i f_i \;\;\; \rho {\bf u}=\sum_i f_i {\bf v}_i.
\end{equation}
To simulate a two-component mixture we now have to consider the
densities of the two fluids $\rho_1$ and $\rho_2$. The first lattice
Boltzmann equation (\ref{eqnLB1}) is now an equation for the total
density $\rho = \rho_1+\rho_2$ and we introduce a second lattice
Boltzmann equation for $g_i$ to describe the evolution of the density
difference $\phi = \rho_1-\rho_2$:
\begin{eqnarray}
&&g_i({\bf x}+{\bf v}_i \Delta t,t+\Delta t)=\nonumber\\
&&g({\bf x},t)
+ \frac{\Delta t}{\tau} \left[g_i^0({\bf x},t)-g_i({\bf x},t)\right]
\end{eqnarray}
where we choose a diagonal collision matrix with a single relaxation
time $\tau$. These densities are only defined on the first nine
velocities ${\bf v}_i$. The density difference $\phi$ is given by
\begin{equation}
\phi = \sum_i g_i.
\end{equation} 
By choosing appropriate equilibrium distributions and an appropriate
collision matrix, we ensure that the following partial differential
equations are being simulated up to second order in the derivatives
but assuming that the relaxation of the viscoelastic stress $\sigma$
is slow ($\theta \sim 1/\sqrt{\epsilon}$):
\begin{eqnarray}
\partial_t \rho + \partial_{\bf x} (\rho {\bf u}) &=& 0\\
\rho \partial_t {\bf u} + \rho {\bf u}.\nabla {\bf u} &=&
-\partial_{\bf x} P + \partial_{\bf x} (\sigma_v+\sigma)\\
\sigma_v &=& \nu_\infty ( \nabla (\rho {\bf u}) + (\nabla
(\rho {\bf u}))^T - \nabla.{\bf u} \delta)\nonumber\\
&&+ \xi_\infty \nabla.{\bf u}\delta\\
\sigma + \theta(\phi) \sigma_{(1)} &=& -
(\nu_0(\phi)-\nu_\infty)(\nabla (\rho {\bf u})+ (\nabla (\rho {\bf u}))^T)\label{jeffrey}\\
\partial_t \phi + \partial_{\bf x}(\phi {\bf
u})&=& D \nabla^2 \mu + \nabla.((\phi/\rho)\nabla.(P-\sigma)) \label{diffusion}
\end{eqnarray}
where $\delta$ is the identity matrix, $\sigma$ is the viscoelastic
stress tensor, $\sigma_{(1)}=\partial_t \sigma + {\bf u}.\nabla
\sigma- \sigma.(\nabla {\bf u})-(\nabla {\bf u})^T \sigma$ is its
upper convected derivative, $P=0.5 \rho+0.007 (\nabla \phi \nabla \phi
- 0.5 \nabla \phi.\nabla \phi \delta- \phi \nabla^2 \phi \delta$ is
the pressure tensor, and $\mu=-0.55 \phi/\rho + 0.25 \ln
((\rho+\phi)/(\rho-\phi)) - 0,007 \nabla^2 \phi$ is the chemical
potential. The parameters $\nu_\infty$, $\xi_\infty$, and $\theta$ are
determined by the eigenvalues of the collision matrix. The values for
the parameters were $\Delta t=1$, $\tau=1$, $D=0.5$, $\xi_\infty=0.31$, and
$\nu_\infty=0.01$. For the low-viscosity phase we used $\theta=0.055$,
$\nu_0=0.013$ and for the viscoelastic phase $\theta=39.5$ and
$\nu_0=1.97$. For the symmetric simulations of Figure \ref{fig:comp}
we used $\theta=0.055$ and $\nu_0\approx \nu_\infty=0.075$.

\section{Simulations}
We performed simulations of critical spinodal decomposition of a
viscoelastic binary mixture in two dimensions where one component is
much more viscoelastic than the other. These simulations lead to the
usual morphologies in which the viscoelastic phase is connected and
the less viscoelastic phase is dispersed\cite{ve-decomp}. We performed
our simulations on a $256^2$ and a $1024^2$ lattice and after about
1000 iterations the less viscoelastic phase is completely dispersed,
although the domains are still highly deformed. We used this
morphology as an initial condition for a simulation where we make both
components purely viscous to examine the effects of initial conditions
that are not symmetric on symmetric binary fluid mixtures. We choose
the viscosity such that a system started with a random initial
condition will be in the viscous scaling state. This allows us to
distinguish the effect that the morphology created by viscoelastic
phase-separation has from the effect of viscoelasticity itself in the
late-time phase-ordering process.

\begin{figure}
\begin{center}
\begin{minipage}{7cm}
\begin{minipage}{3cm}
\centerline{\psfig{figure=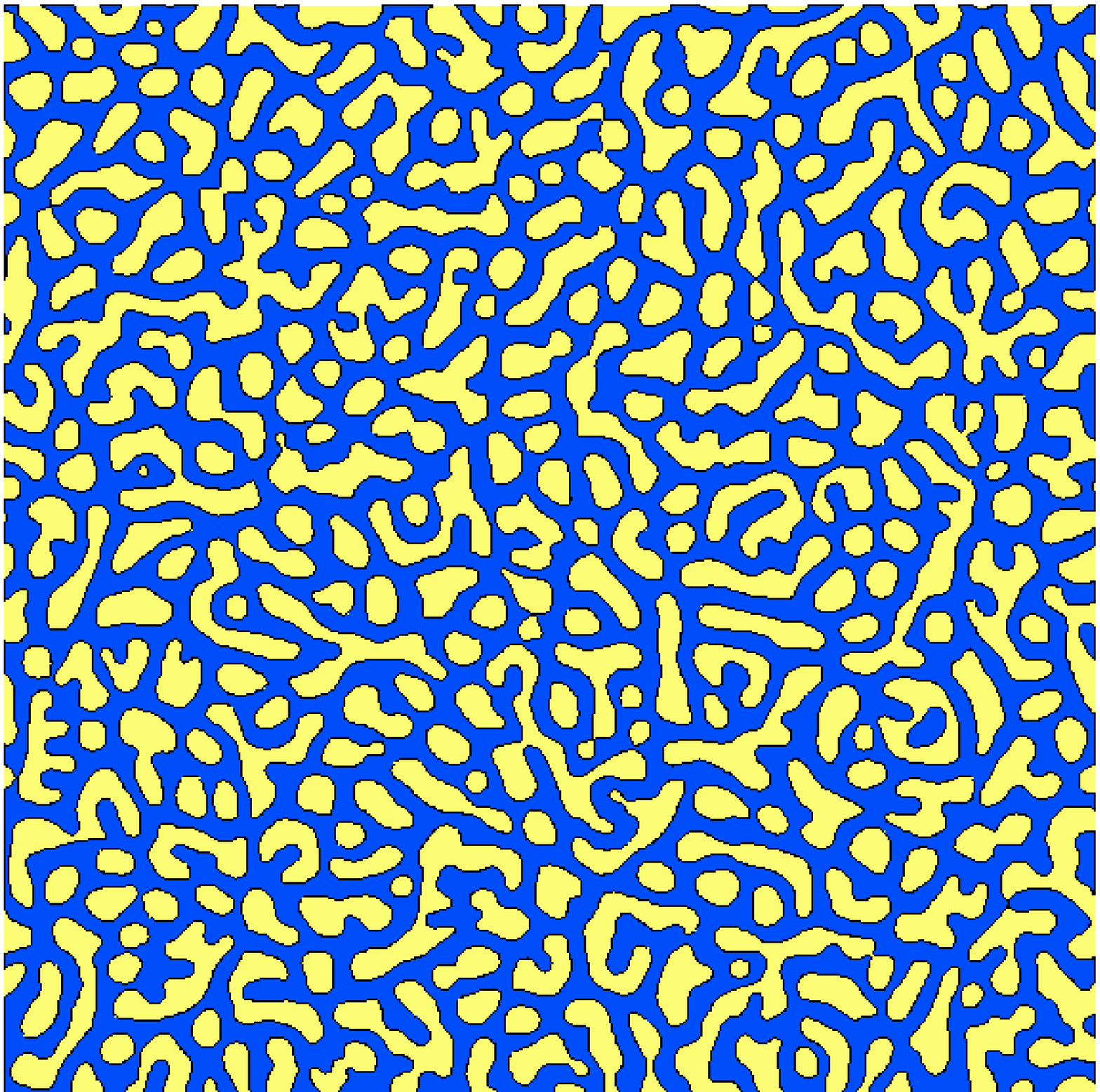,width=2.8cm}}
\end{minipage}
\hspace{0.3cm}
\begin{minipage}{3cm}
\centerline{\psfig{figure=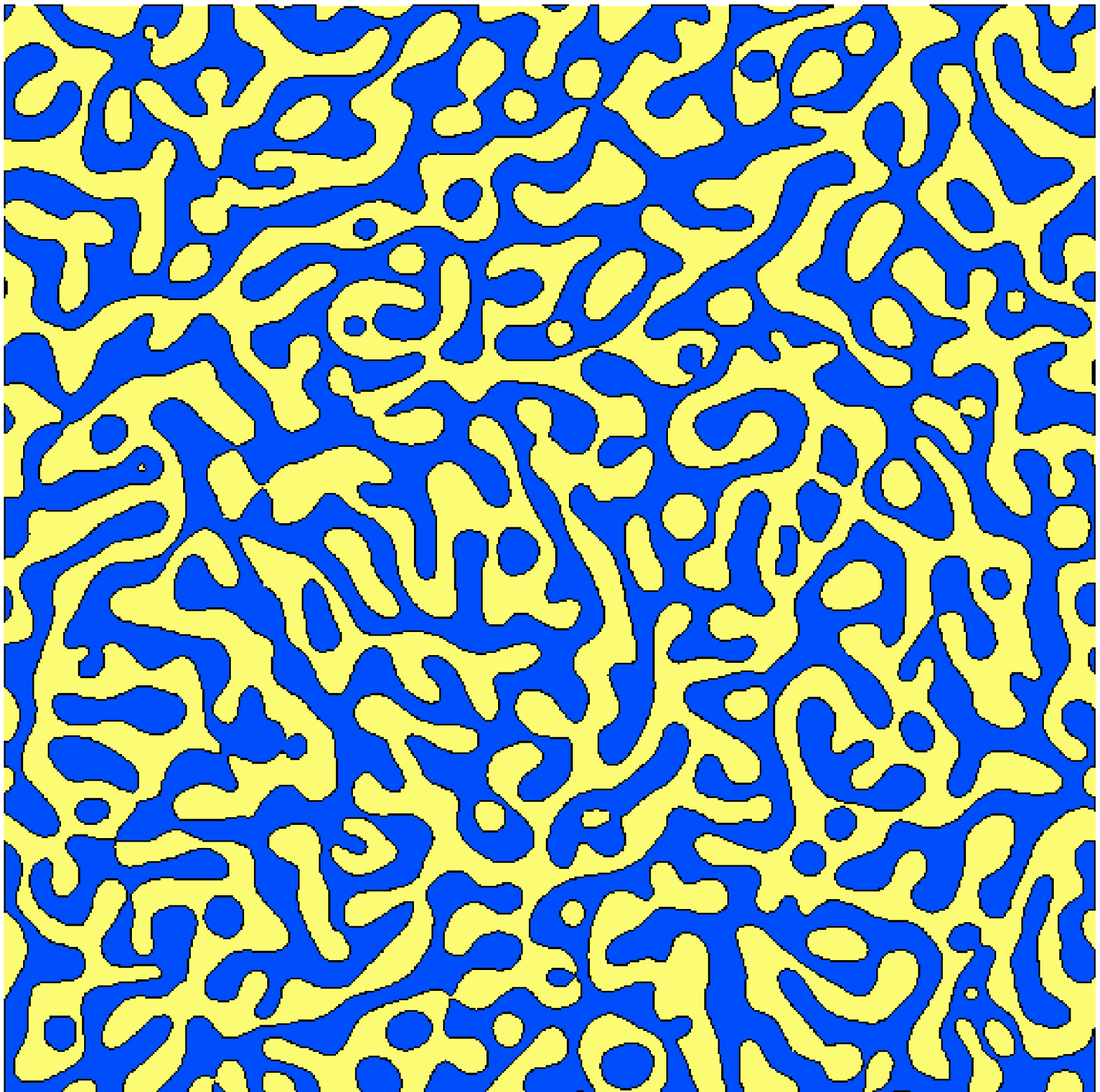,width=2.8cm}}
\end{minipage}\\
\begin{minipage}{3cm}
\centerline{\psfig{figure=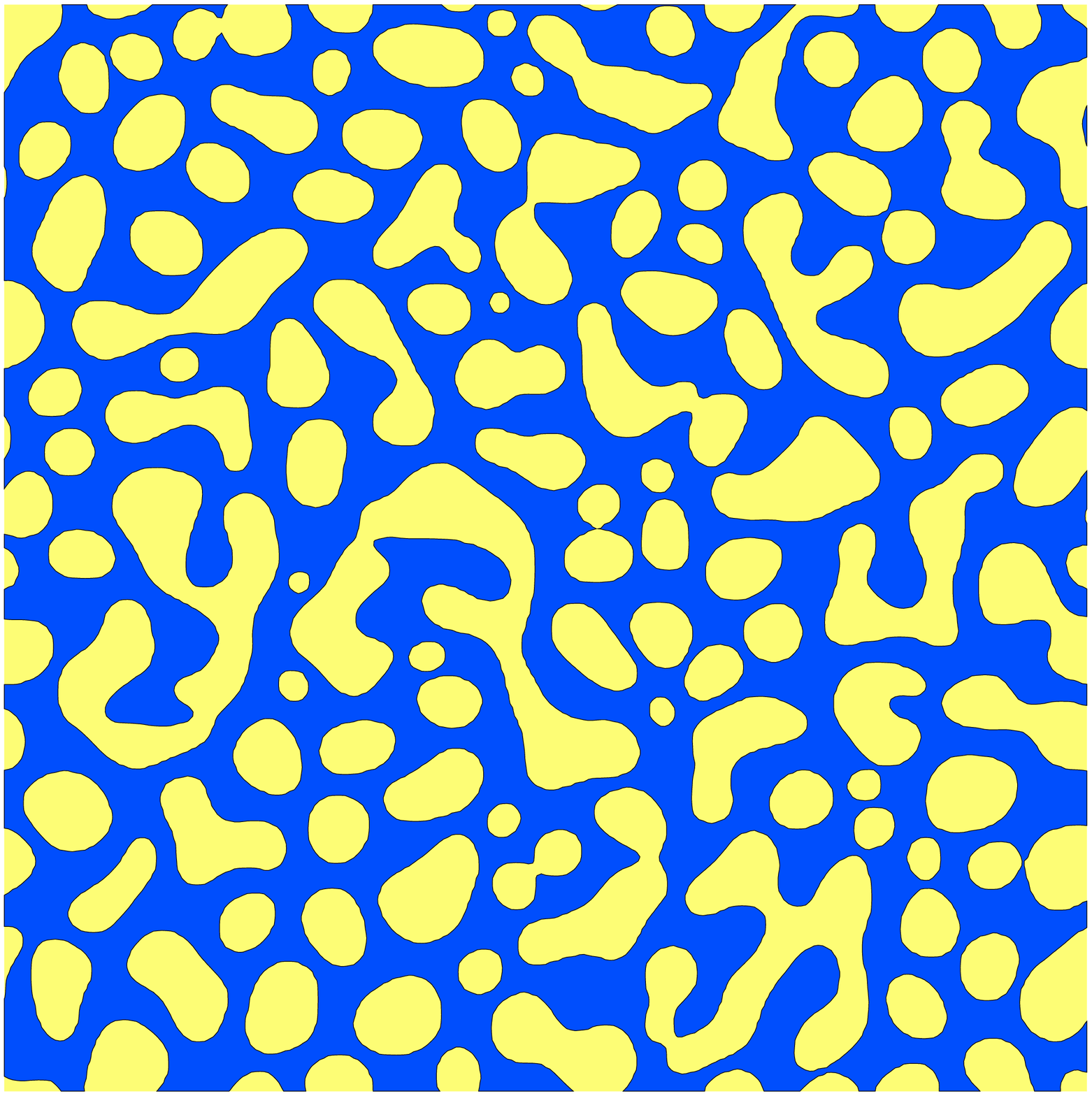,width=2.8cm}}
\end{minipage}
\hspace{0.3cm}
\begin{minipage}{3cm}
\centerline{\psfig{figure=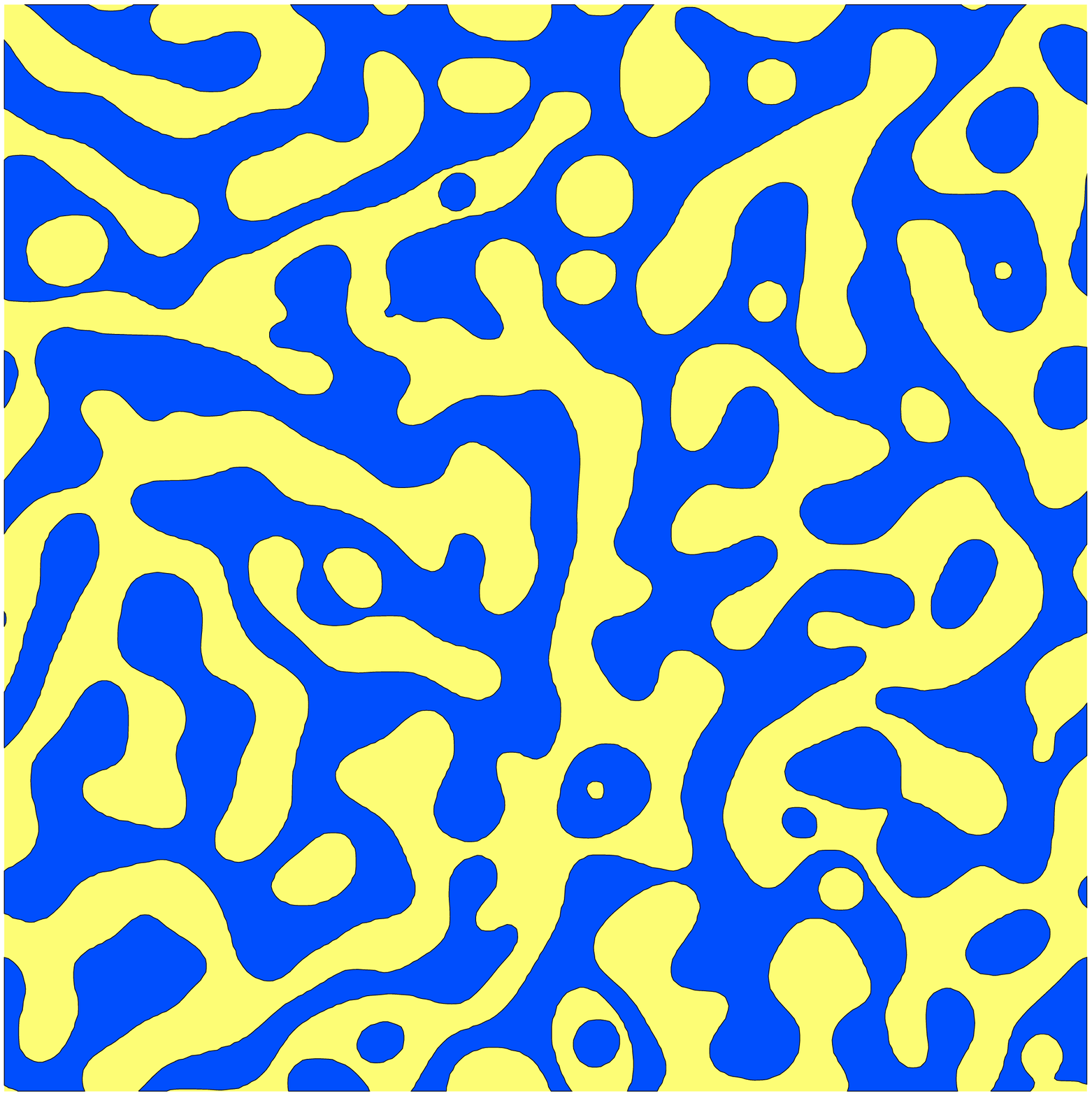,width=2.8cm}}
\end{minipage}\\
\begin{minipage}{3cm}
\centerline{\psfig{figure=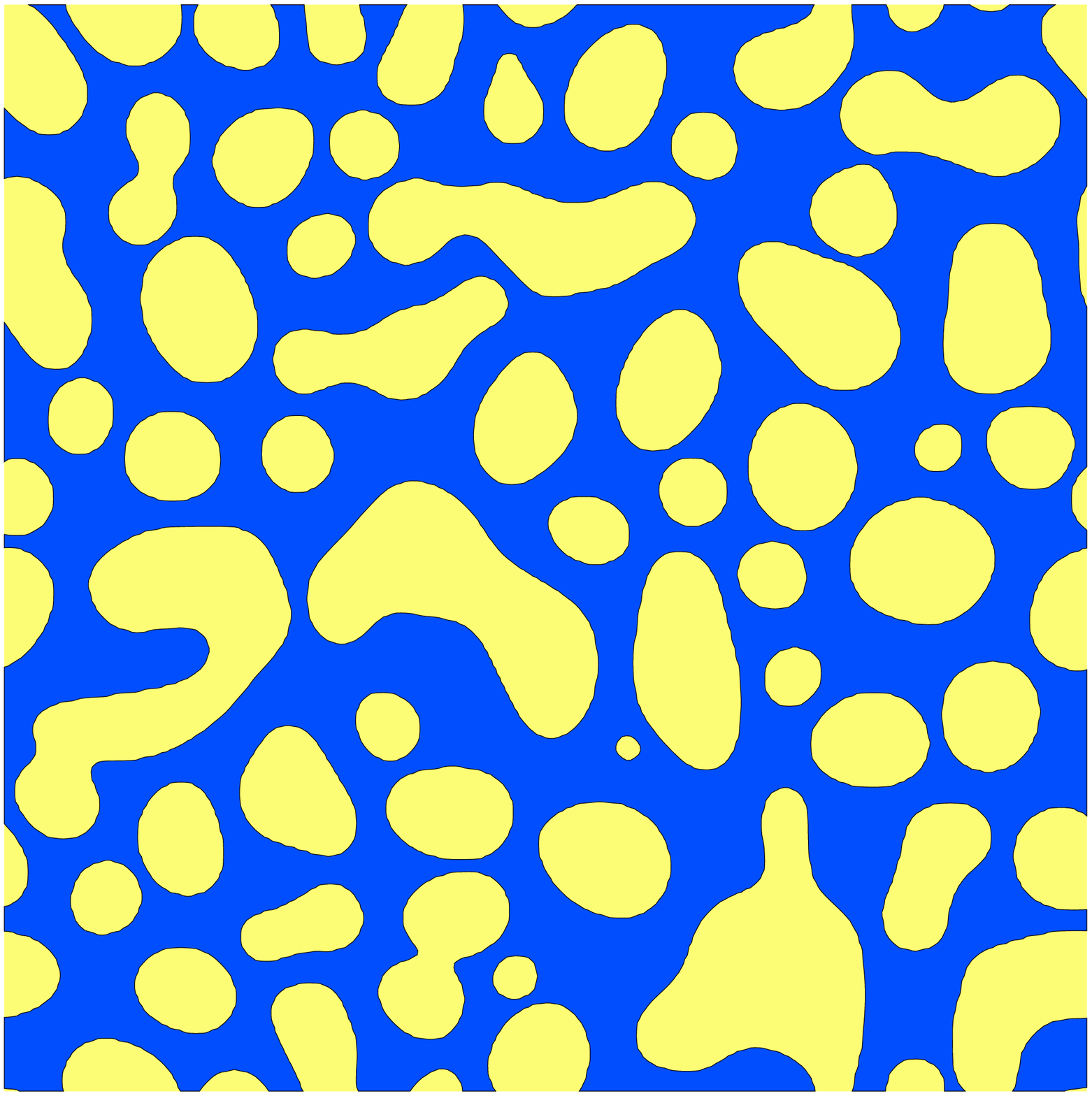,width=2.8cm}}
\end{minipage}
\hspace{0.3cm}
\begin{minipage}{3cm}
\centerline{\psfig{figure=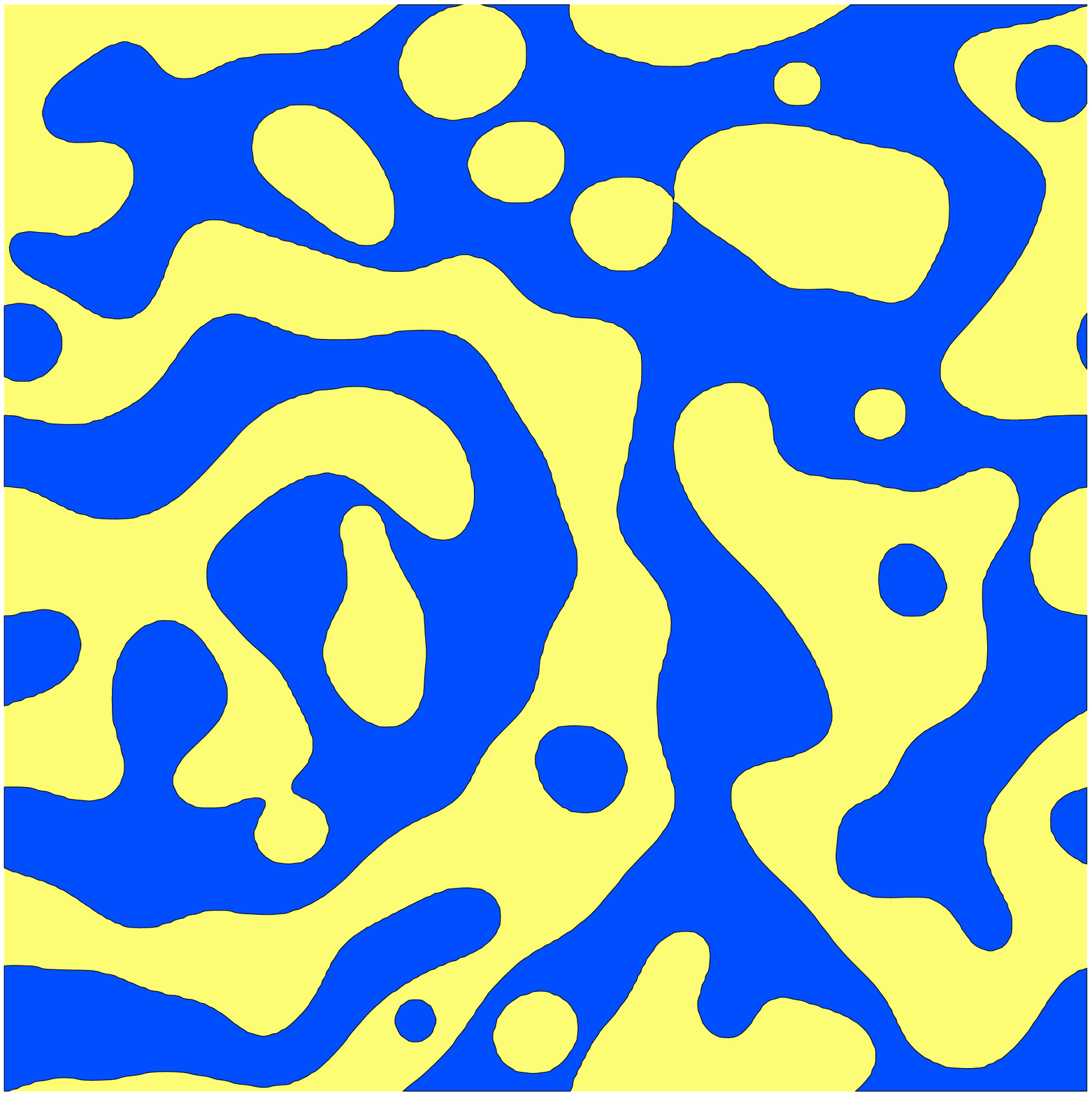,width=2.8cm}}
\end{minipage}\\
\begin{minipage}{3cm}
\centerline{\psfig{figure=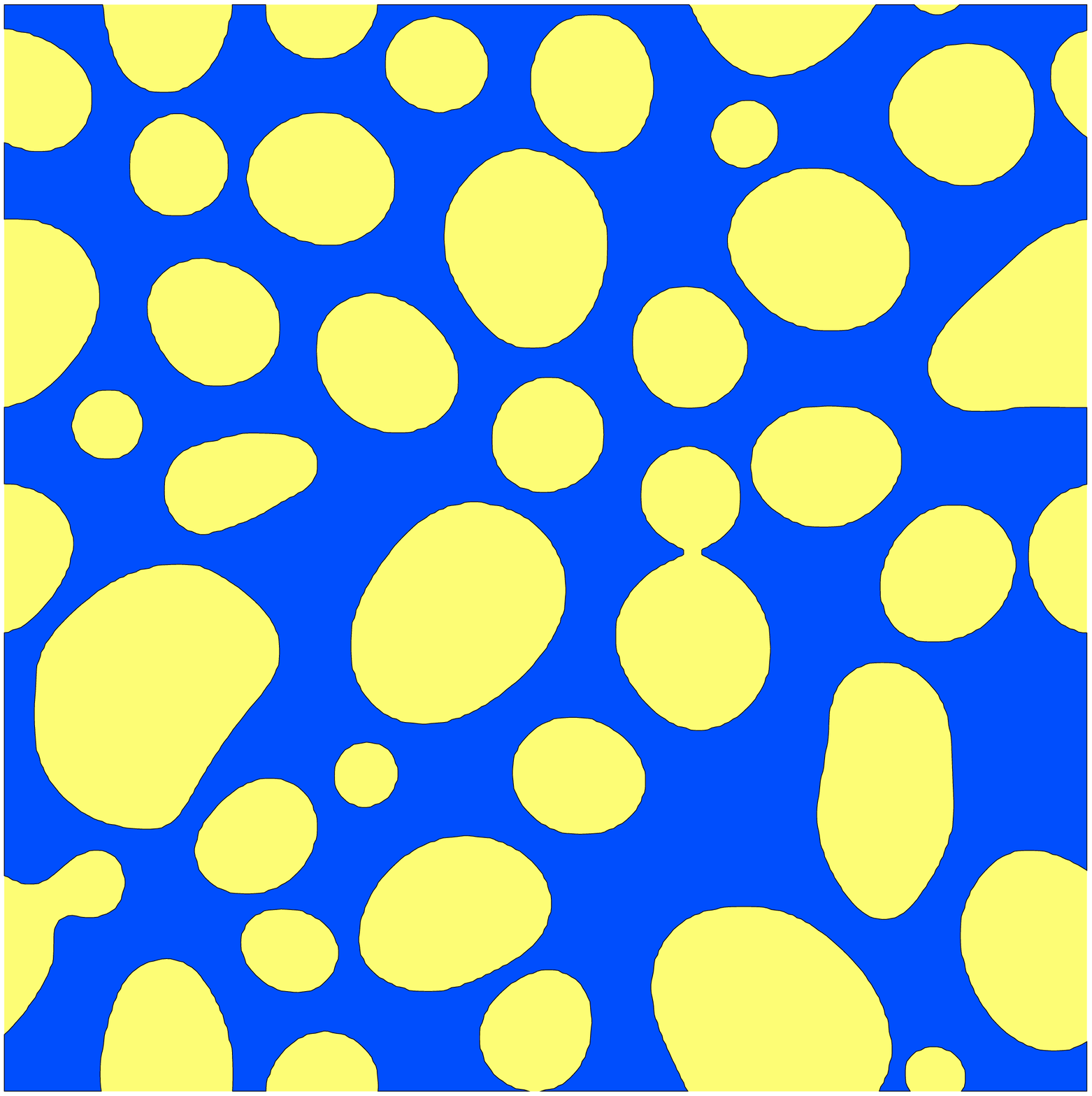,width=2.8cm}}
\end{minipage}
\hspace{0.3cm}
\begin{minipage}{3cm}
\centerline{\psfig{figure=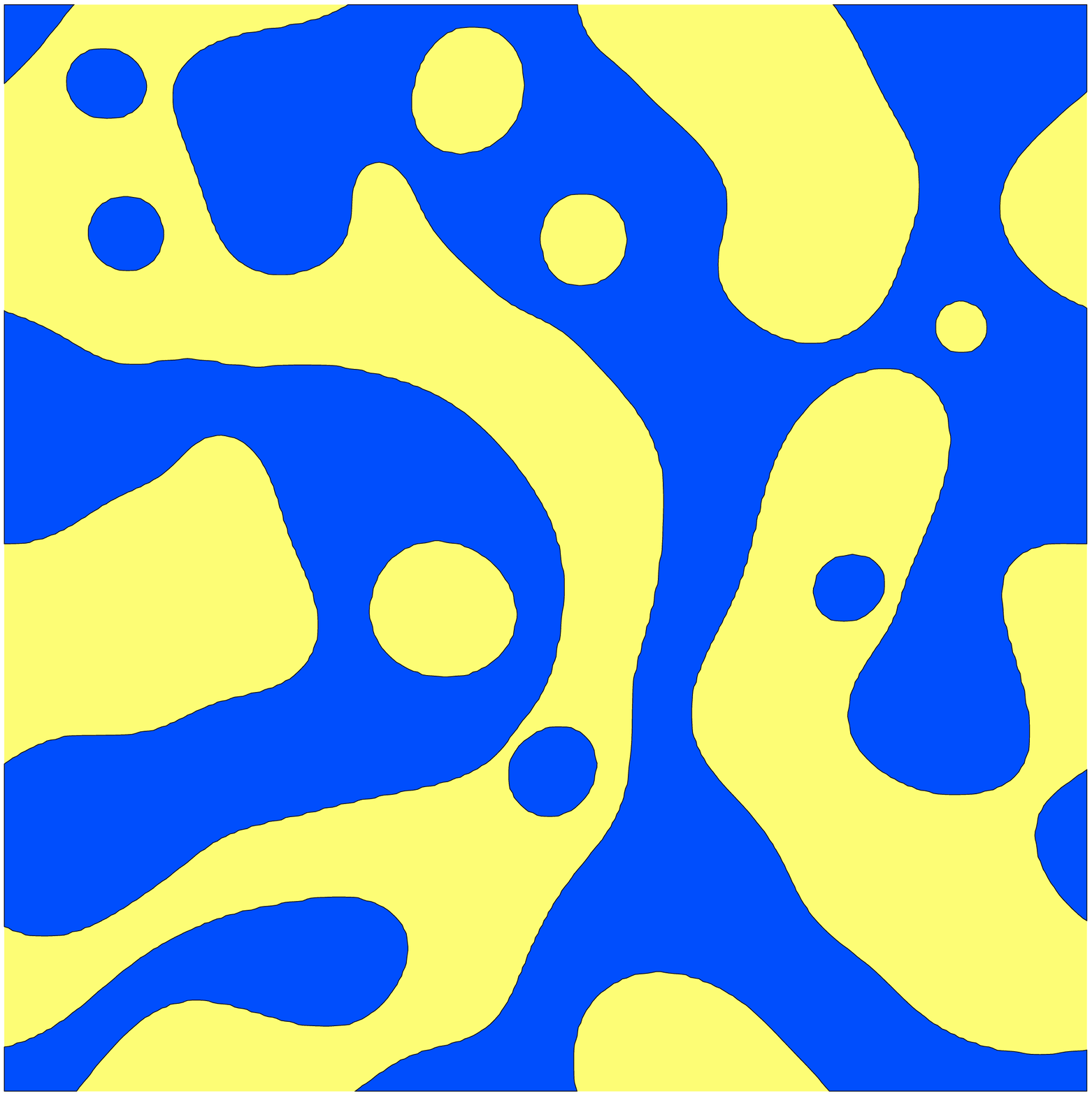,width=2.8cm}}
\end{minipage}
\begin{center}
(a) \hspace{2.5cm} (b) 
\end{center}
\begin{minipage}{6cm}
\centerline{\psfig{figure=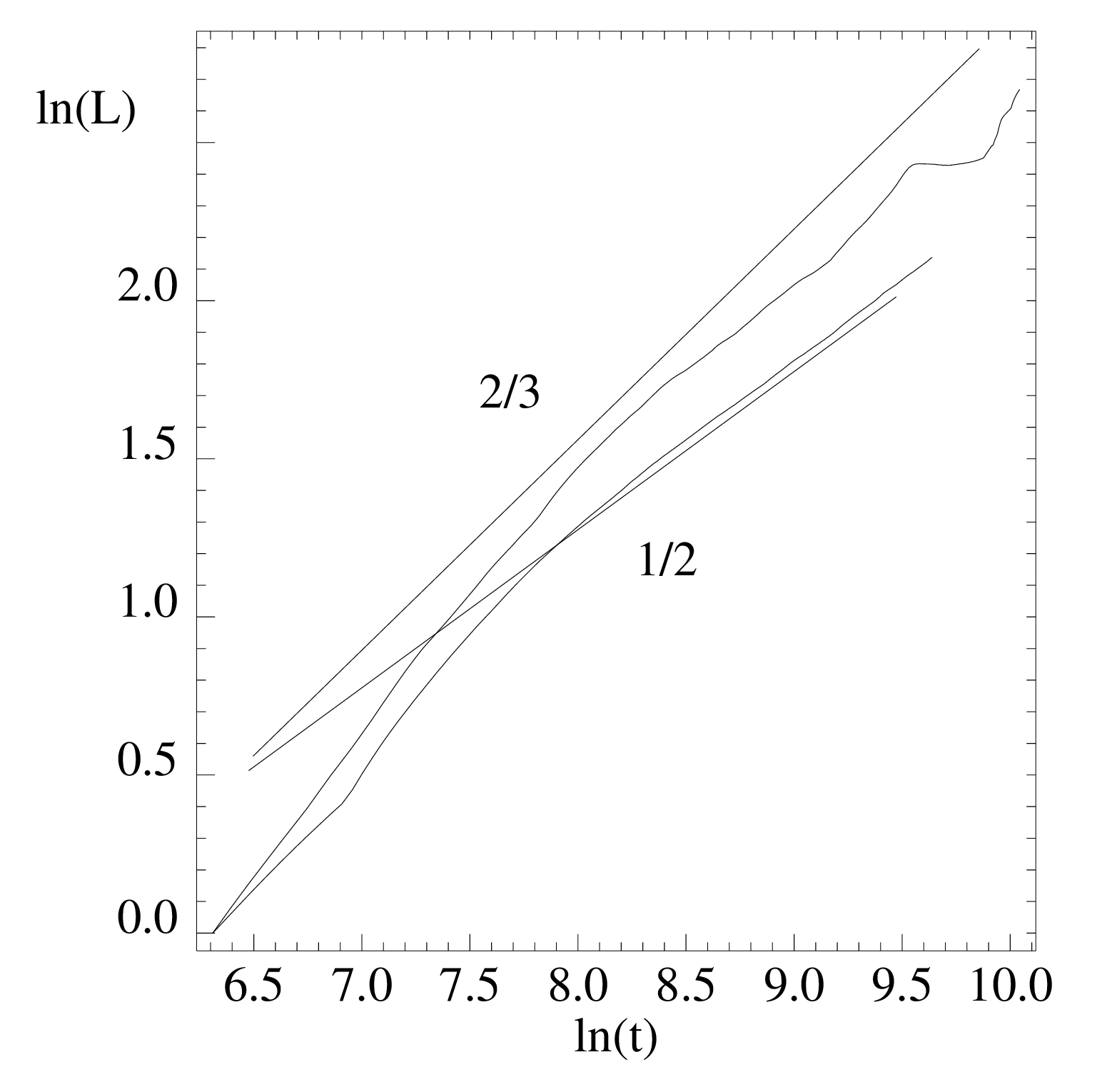,width=5cm}}
\end{minipage}
\begin{center}
(c)
\end{center}
\end{minipage}
\end{center}
\caption{A comparison of the phase-ordering of two identical symmetric
binary mixtures after different early time spinodal decomposition. In
(a) a phase-ordering is seen from an initial morphology generated by
viscoelastic phase-separation in which the light component is
dispersed (the originally low-viscous component). In (b) we see the
usual symmetric phase-separation. Fig.(c) shows the length scales
$L(t)$ for (a) $L\sim t^{1/2}$ and (b) $L\sim t^{2/3}$. The apparent
growth law in (a) is misleading. A more detailed study that examines
the growth of the droplet morphology over several decades in length
shows that the growth-law is $L\sim t$ (see \cite{AaC1}).}
\label{fig:comp}
\end{figure}

In Figure \ref{fig:comp} we see a comparison of the phase-ordering
from a droplet morphology (a) and a symmetric initial condition
(b). The morphologies for both systems are shown after 1000, 2000,
4000, and 8000 iterations. We see that the phase-ordering of the
morphology with a dispersed phase leads to an even more pronounced
dispersed phase where almost all domains are circular at late
times. We see that droplet coalescence occurs frequently with only few
domains vanishing due to the evaporation-condensation mechanism
underlying Oswald ripening. The droplet coalescence, however, is not
frequent enough to change the connectivity of the domains. Instead, we
observe that domains become more circular on average, suggesting that
even for very long times we do not expect a transition to a
bi-continuous morphology. (This is no longer true when we leave
viscous regime and enter into the inertial regime. Here a return to
the inertial scaling state is observed\cite{AaC1}). In Figure
\ref{fig:comp}(c) we see that the growth law for the droplet phase
appears to be $L(t) \sim t^{1/2}$ and is smaller than the $L(t)\sim
t^{2/3}$ seen for the symmetric phase-ordering shown in Figure
\ref{fig:comp}(b). More recent simulations of the droplet morphology,
however, establish scaling over several orders of magnitude and find
that the actual growth-law is $L\sim t$ \cite{AaC1}. This emphasizes
the fact that dynamic scaling analysis that does not cover several
decades can be misleading. Unfortunately such studies are very
computationally demanding and exist only in few studies.

The existence of this second scaling state, distinct from the
bi-continuous scaling state, is important to understanding the
late-time regime of viscoelastic phase-separation because, even in the
absence of viscoelastic effects, we observe a droplet-morphology
evolving from the initial morphology created by viscoelastic
phase-separation. This result is also important for practical reasons
in processes where late-time morphologies need to be controlled. It is
well known that mixing of high-viscosity and low-viscosity components
by means of mechanical agitation leads to morphologies where the
high-viscosity phase is dispersed in droplets\cite{Utracki} for volume
fractions of the low-viscosity component of much less than 50\%. This
effect is enhanced if the high-viscous phase is
viscoelastic\cite{Vanoene}. This leads us to consider what the
late-time morphology of a phase-ordering system with an early
morphology created by mechanical mixing would be.

\begin{figure}
\begin{center}
\begin{minipage}{7cm}
\begin{minipage}{3cm}
\centerline{\psfig{figure=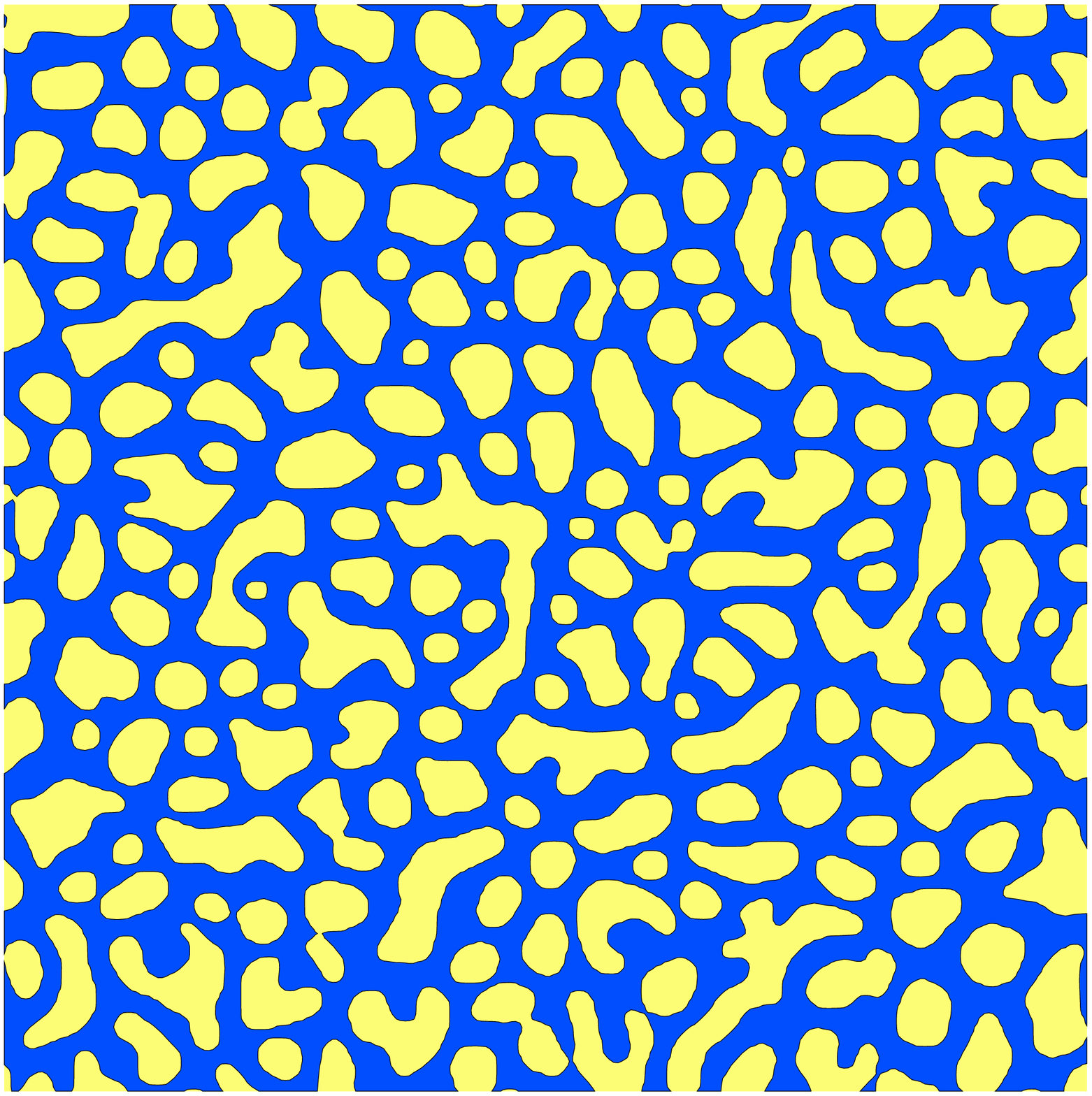,width=2.8cm}}
\end{minipage}
\hspace{0.3cm}
\begin{minipage}{3cm}
\centerline{\psfig{figure=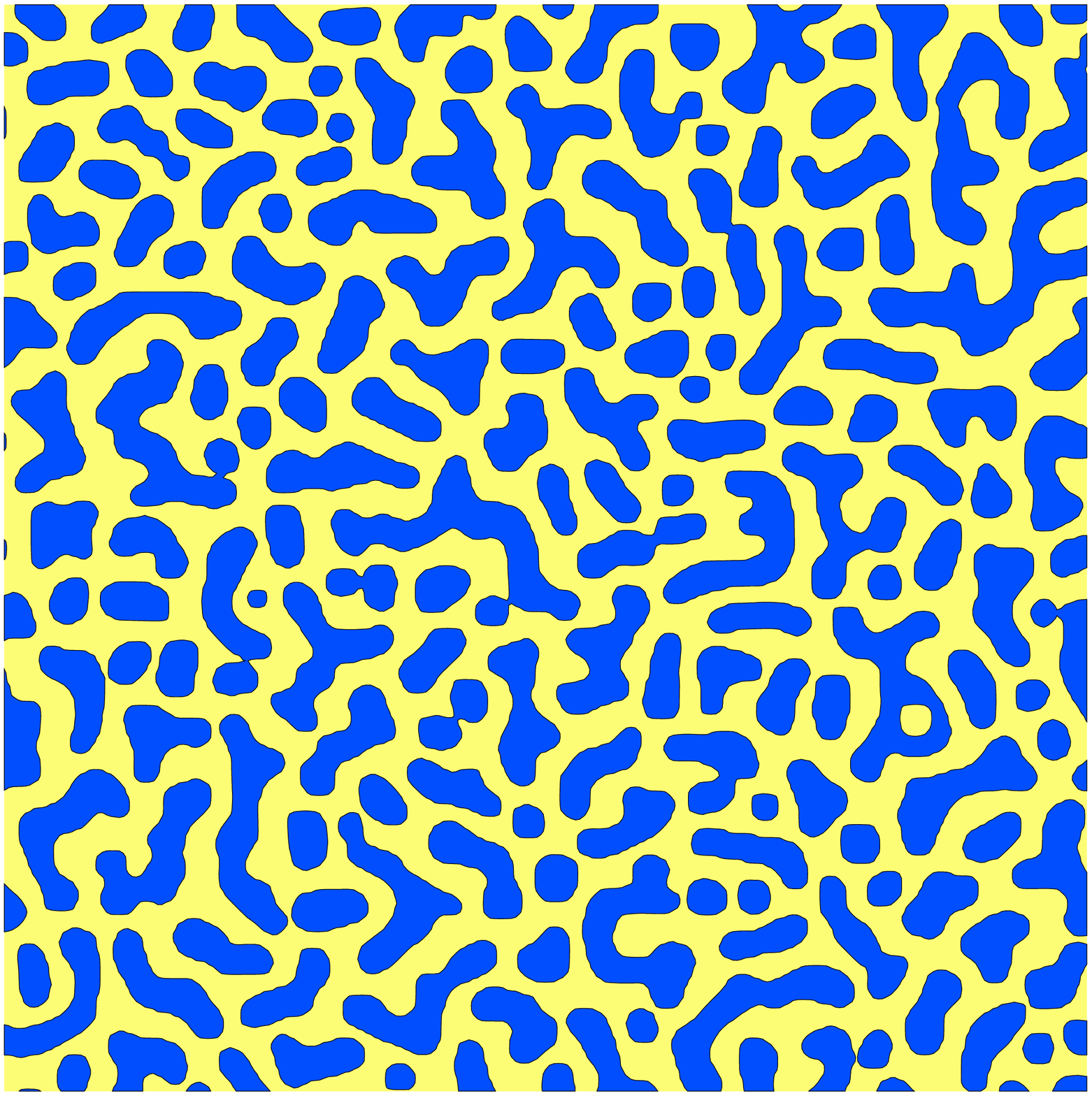,width=2.8cm}}
\end{minipage}\\
\begin{minipage}{3cm}
\centerline{\psfig{figure=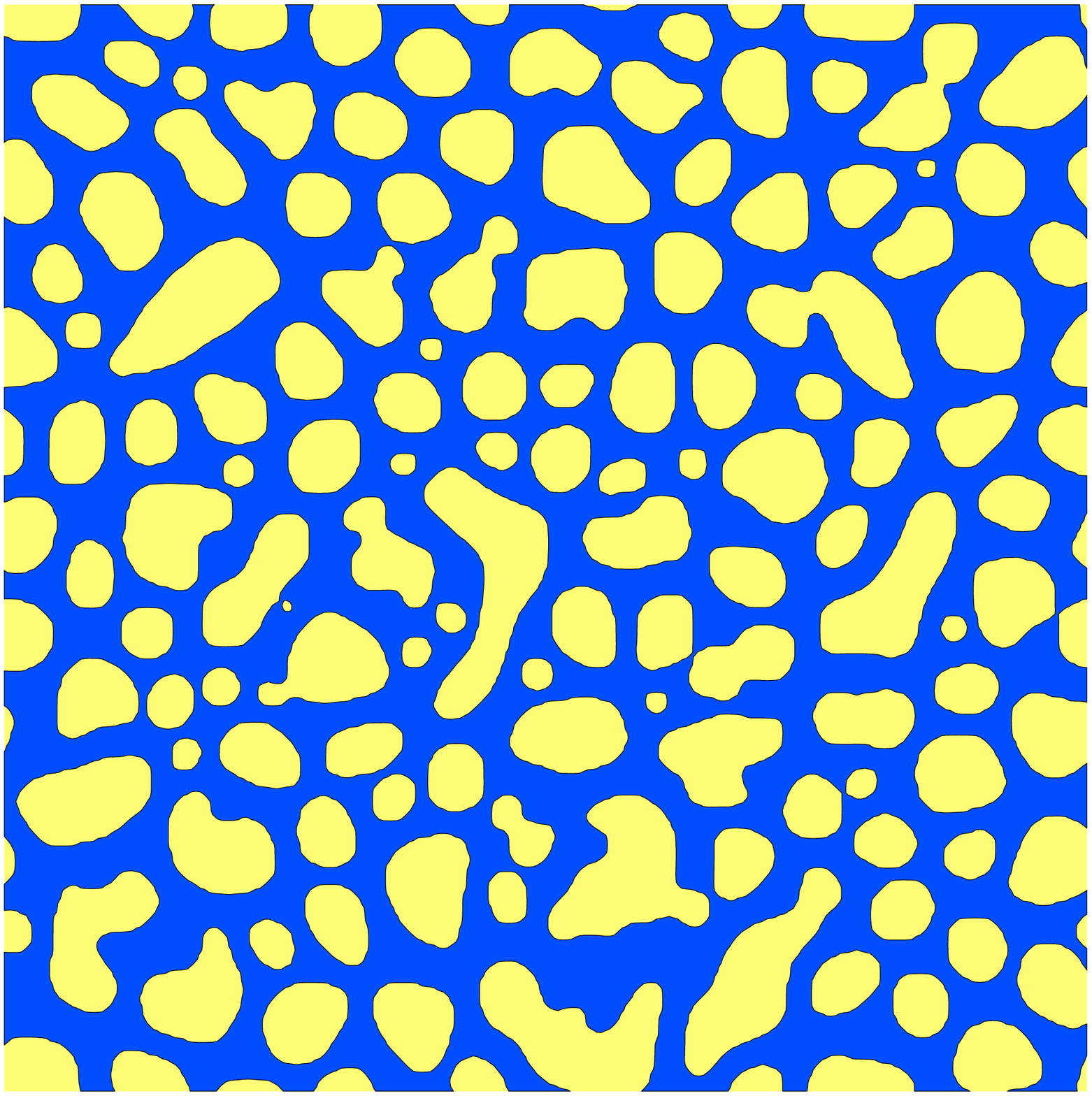,width=2.8cm}}
\end{minipage}
\hspace{0.3cm}
\begin{minipage}{3cm}
\centerline{\psfig{figure=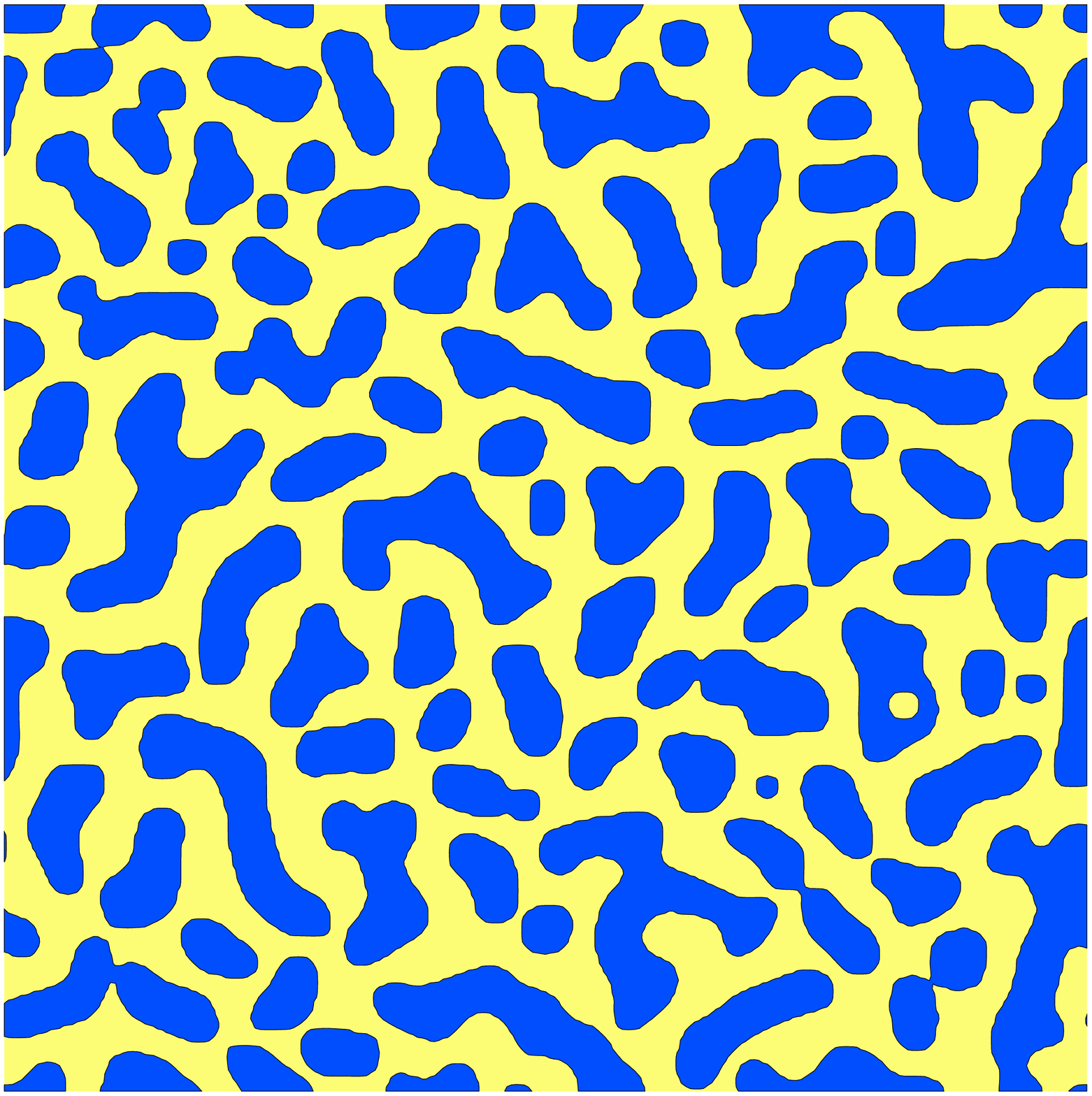,width=2.8cm}}
\end{minipage}\\
\begin{minipage}{3cm}
\centerline{\psfig{figure=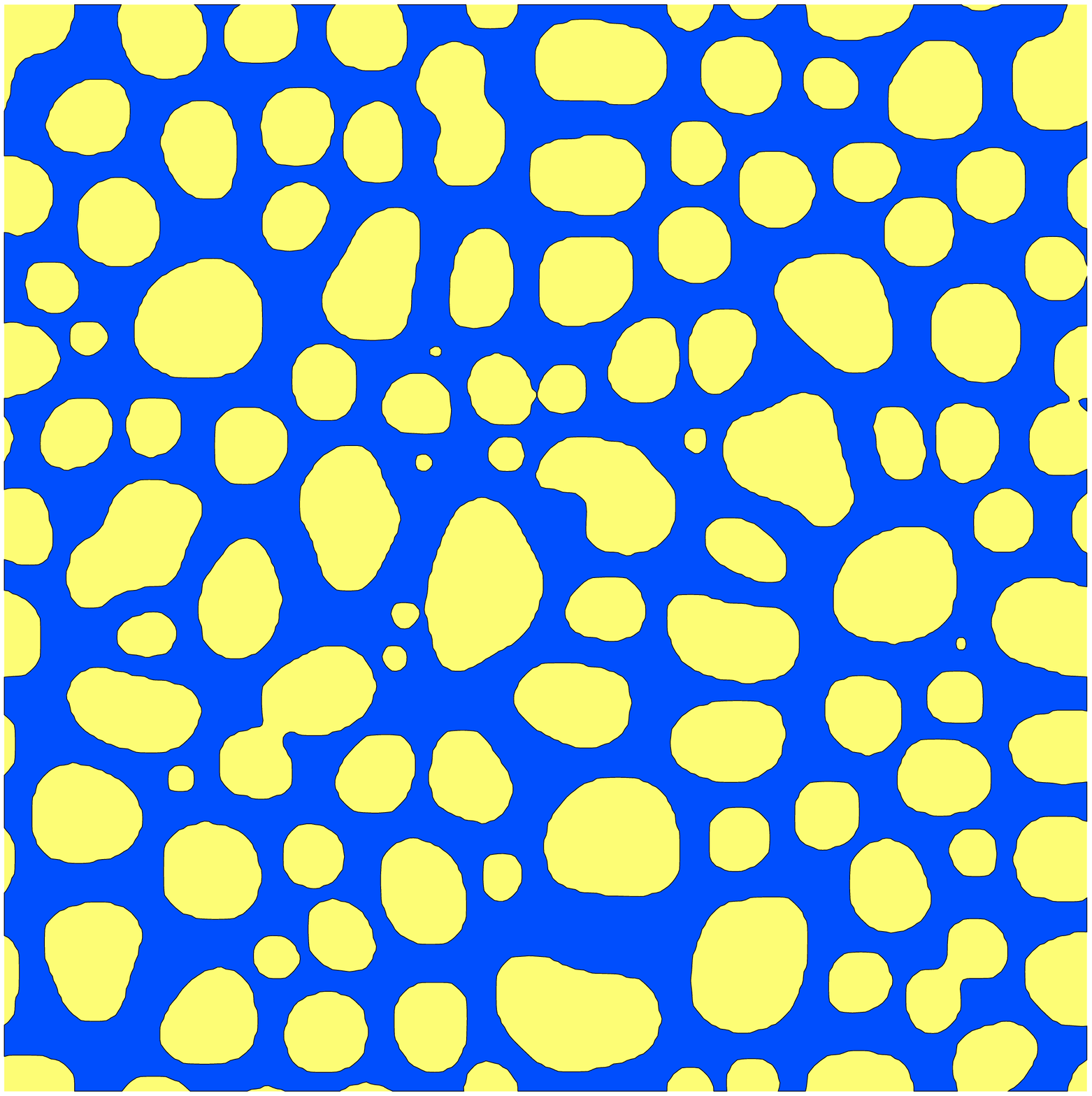,width=2.8cm}}
\end{minipage}
\hspace{0.3cm}
\begin{minipage}{3cm}
\centerline{\psfig{figure=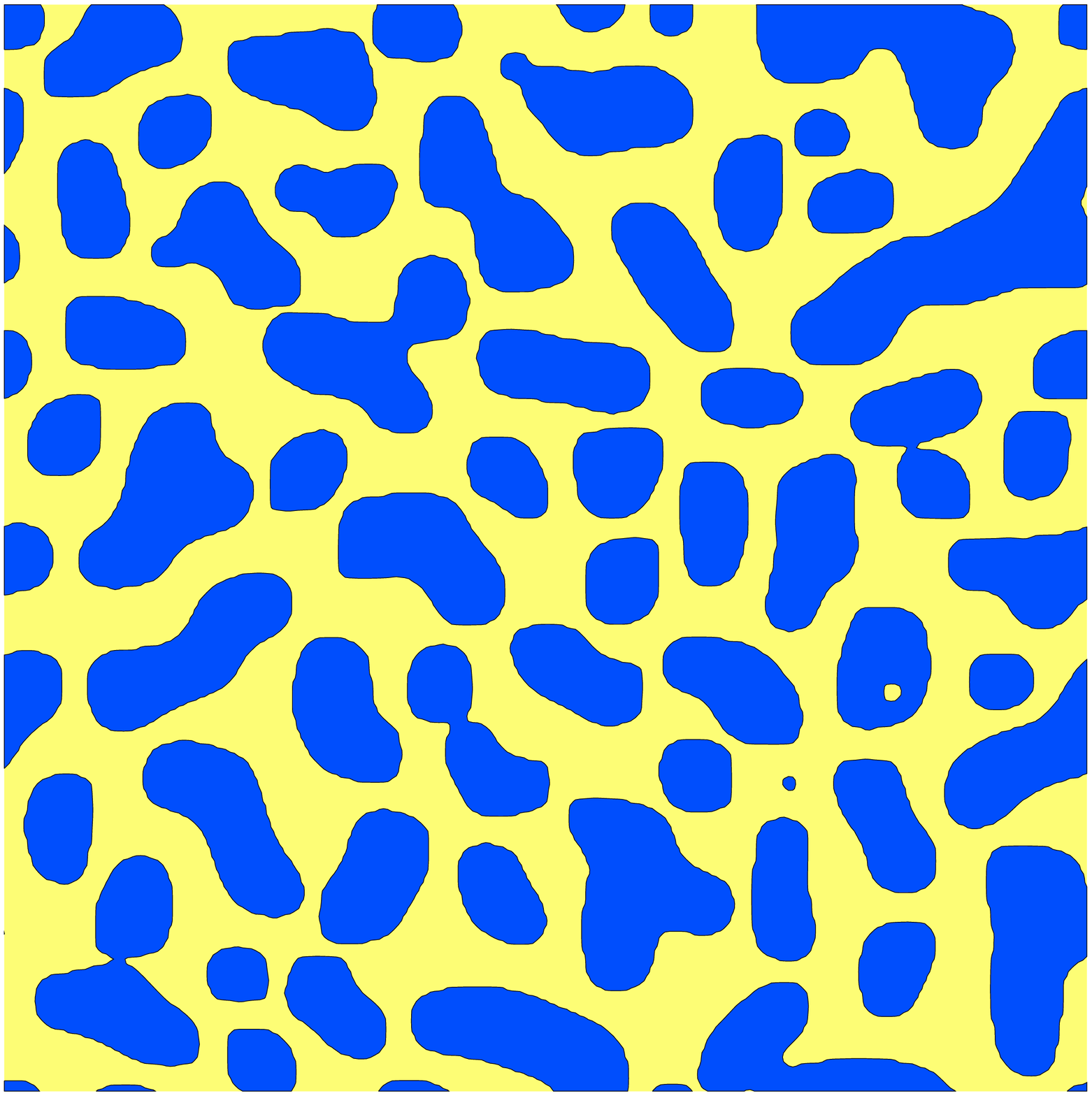,width=2.8cm}}
\end{minipage}\\
\begin{minipage}{3cm}
\centerline{\psfig{figure=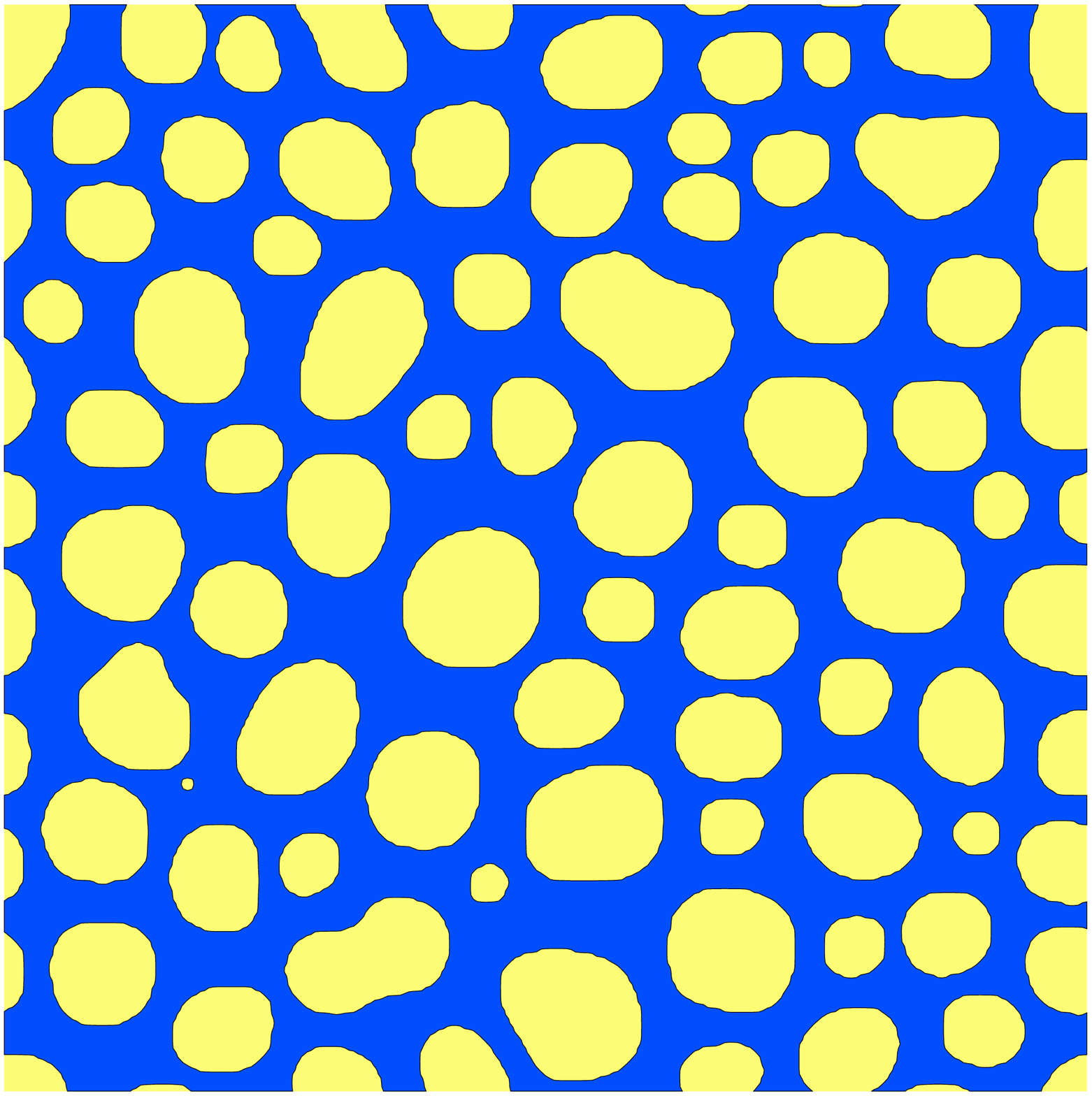,width=2.8cm}}
\end{minipage}
\hspace{0.3cm}
\begin{minipage}{3cm}
\centerline{\psfig{figure=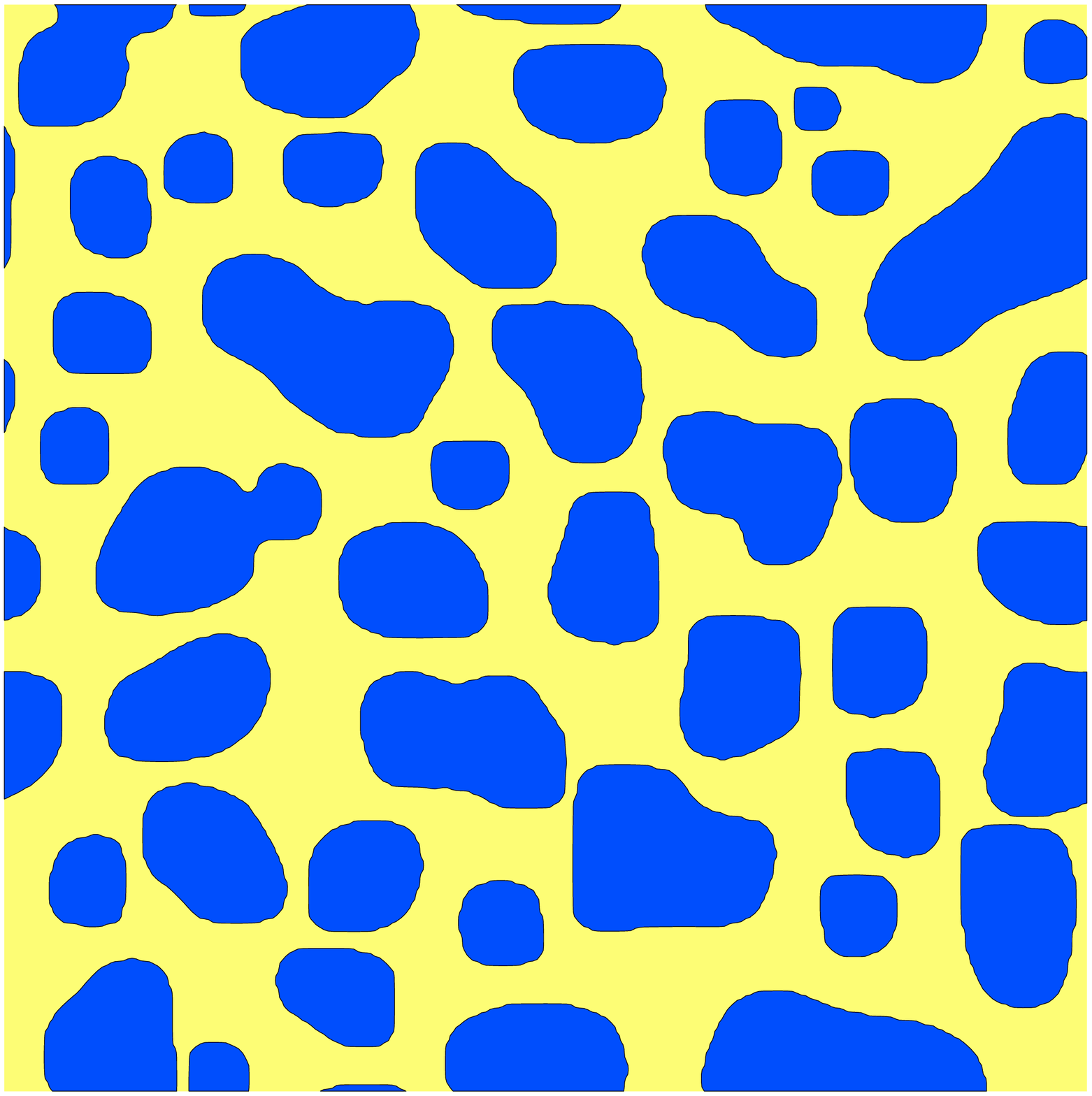,width=2.8cm}}
\end{minipage}
\begin{center}
(a) \hspace{2.5cm} (b) 
\end{center}
\begin{minipage}{6cm}
\centerline{\psfig{figure=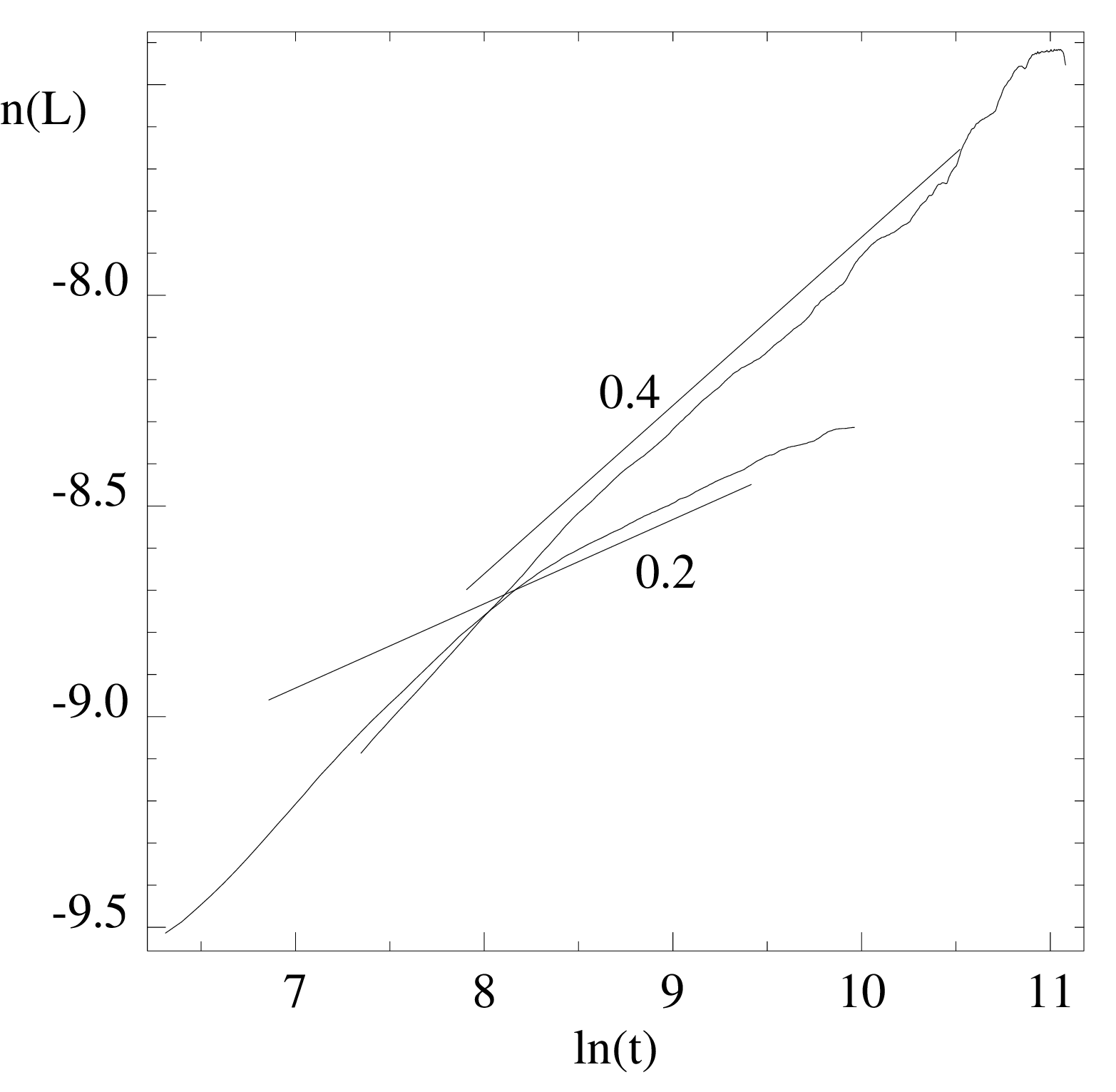,width=5cm}}
\end{minipage}
\begin{center}
(c)
\end{center}
\end{minipage}
\end{center}
\caption{The phase-ordering of a binary mixture of a viscoelastic
phase (dark color) and a low viscosity Newtonian phase (light color)
is shown in (a) and (b). (a) corresponds to a usual viscoelastic
phase-separation of a 50\%-50\% mixture and (b) shows the
phase-ordering after the viscoelastic and low-viscosity domains have
been interchanged. Morphologies are shown after 2000, 4000, 8000, and
16000 iterations. The scaling behavior of the two mixtures is shown in
(c) where for (a) $L\sim t^{0.2}$ and for (b) $L\sim 0.4$. As in
Fig. \ref{fig:comp} these power-laws have not been established for
several decades and should therefore not be taken too literally.}
\label{fig:vecomp}
\end{figure}

In order to answer this question we performed a simulation of
viscoelastic phase-separation and after a droplet morphology had been
formed at 1000 time-steps, we inverted the properties of the two
components. We used this state as a model for the morphologies of
dispersed droplets of the viscoelastic phase in a matrix of the low
viscosity Newtonian phase that is typical of mechanically mixed
morphologies. We then continued the simulations and observed the
phase-ordering behavior of the new morphology. The results of these
simulations are shown in Figure \ref{fig:vecomp}. In Figure
\ref{fig:vecomp}(a) the morphology for a viscoelastic phase-separation
of a 50\%-50\% mixture is shown. We should emphasize that the
phase-ordering (see eqn. (\ref{diffusion})) does not have a
composition-dependent diffusion constant and therefore no domain
shrinkage is observed in these simulations. Domain shrinkage can be a
very slow process in viscoelastic phase-separation that prolongs the
spinodal decomposition process and makes it more difficult to
differentiate the early-stage decomposition and the late-time
phase-ordering processes. This does not reduce the validity of our
results, however, since we are only interested in the late-time
behavior when the domain shrinkage is completed.

The morphologies shown in Figure \ref{fig:vecomp}(b) are of a
simulation where at 1000 iterations, after the dispersion was
achieved, the viscoelastic and the low viscosity components were
exchanged. We see that the morphology remains a droplet morphology,
albeit now of the viscoelastic phase. Comparing Figures
\ref{fig:vecomp} (a) and (b) we see that the most important factor in
determining the final morphology is the early stage dynamics, and that
phase-ordering scaling morphologies of both dispersed viscoelastic
and dispersed low viscosity domains exist. 

From Figure \ref{fig:vecomp}(c) we see that each scaling state appears
to have a different growth law. The morphology of dispersed
viscoelastic domains grows as $L\sim t^{0.4}$ whereas the morphology
of dispersed low viscosity domains grows as $\sim t^{0.2}$ after near
circular droplets have been formed. An anomalous slow growth in
viscoelastic phase-separation has first been observed experimentally
by Tanaka\cite{tanaka} who found $L\sim t^{0.15}$ for a system of high
molecular-weight polystyrene/ diethyl malonate (4.0 wt.\%). It must,
however, be emphasized that it is impossible to conclusively determine
scaling exponents from such a short run. It is conceivable that a more
extensive scaling analysis could lead to different exponents.

These simulations also emphasize the difference between a morphology
after spinodal decomposition and after mechanical mixing. The
morphology after spinodal decomposition is a dispersed low viscosity
phase, whereas the state after mechanical mixing has a dispersed
viscoelastic phase. Subsequent phase-ordering does not change the
connectivity of these states, in agreement with the conventional
wisdom that there is a profound difference in states produced by
spinodal decomposition and mechanical mixing.

\section{Conclusions}
In this article we have shown that more than one scaling state exists
for late-time spinodal decomposition of two-dimensional binary fluids
and that local correlations in the initial conditions or the early
time behavior of a phase-separating binary mixture can be very
important in selecting one of these scaling states. We have also
explained the difference between viscoelastic phase-ordering states
after spinodal decomposition and mechanical mixing. Our results show
that the volume fraction and the physical properties of a mixture do
not select a morphology by themselves, but that the morphology of the
initial state is of paramount importance. This is why viscoelastic
phase-separation can lead to unusual late-time scaling states even
when viscoelasticity is no longer important at large length scales.

\section*{Acknowledgments}
The authors acknowledge the financial support of DuPont Chemical
Company. A.W. acknowledges the support of EPSRC Grant GR/M56234 and
would like to thank Craig Carter for the generous permission to use
his Origin2000 computer.

\def\jour#1#2#3#4{{#1} {\bf #2}, #3 (#4).}
\def\tbp#1{{\em #1}, to be published.}
\def\inpr#1{{\em #1}, in preparation.}
\def\tit#1#2#3#4#5{{#1} {\bf #2}, #3 (#4).}

\def\ap{Adv. Phys.}
\def\arf{Ann. Rev. Fluid Mech.}
\def\epl{Euro. Phys. Lett.}
\def\ijmp{Int. J. Mod. Phys. C}
\def\jcp{J. Chem. Phys.}
\def\jpc{J. Phys. C}
\def\jpcs{J. Phys. Chem. Solids}
\def\jpco{J. Phys. Cond. Mat}
\def\jsp{J. Stat. Phys.}
\def\jf{J. Fluids}
\def\jfm{J. Fluid Mech.}
\def\jnnfm{J. Non-Newtonian Fluid Mech.}
\def\pfa{Phys. Fluids A}
\def\prl{Phys. Rev. Lett.}
\def\pr{Phys. Rev.}
\def\pra{Phys. Rev. A}
\def\prb{Phys. Rev. B}
\def\pre{Phys. Rev. E}
\def\pa{Physica A}
\def\pla{Phys. Lett. A}
\def\ps{Physica Scripta}
\def\roy{Proc. Roy. Soc.}
\def\rmp{Rev. Mod. Phys.}
\def\zpb{Z. Phys. B}


\begin{thebibliography}{200}
\bibitem{Epstein} Irv Epstein showed examples of oscillating chemical
reactions during his presentation at the ICCS3 that clearly exhibited
some phase-ordering behavior.
\end{thebibliography}
\end{document}